\documentclass[twocolumn,showpacs,preprintnumbers,amsmath,amssymb,superscriptaddress,nofootinbib,english]{revtex4-1}
\usepackage{graphicx}
\usepackage{dcolumn}
\usepackage{bm}
\usepackage{epsfig}
\usepackage{graphicx}
\usepackage{hyperref}
\usepackage[usenames]{color}
\usepackage{url}
\usepackage{epstopdf}

\hypersetup{
    colorlinks=true,
    linkcolor=green,
    citecolor=blue,
}

\begin{document}

\title{Large-scale Structure in $f(T)$ Gravity}

\author{Baojiu~Li}
\email[Email address: ]{b.li@damtp.cam.ac.uk}
\affiliation{DAMTP, Centre for Mathematical Sciences, University of Cambridge, Wilberforce Road, Cambridge CB3 0WA, UK}
\affiliation{Kavli Institute for Cosmology Cambridge, Madingley Road, Cambridge CB3 0HA, UK}

\author{Thomas~P.~Sotiriou}
\email[Email address: ]{t.sotiriou@damtp.cam.ac.uk}
\affiliation{DAMTP, Centre for Mathematical Sciences, University of Cambridge, Wilberforce Road, Cambridge CB3 0WA, UK}

\author{John~D.~Barrow}
\email[Email address: ]{j.d.barrow@damtp.cam.ac.uk}
\affiliation{DAMTP, Centre for Mathematical Sciences, University of Cambridge, Wilberforce Road, Cambridge CB3 0WA, UK}

\date{\today}

\begin{abstract}
In this work we study the cosmology of the general $f(T)$ gravity theory. We 
express the modified Einstein equations using covariant quantities, and derive the gauge-invariant perturbation 
equations in covariant form. We consider a specific choice of 
$f(T)$, designed to explain the observed late-time accelerating cosmic expansion without including an 
exotic dark energy component. Our numerical solution shows that the extra degree of freedom of such
$f(T)$ gravity models generally decays as one goes to smaller scales, and consequently its effects on scales such as galaxies
and galaxies clusters are small. But on large scales, this degree of freedom can produce large 
deviations from the standard $\Lambda$CDM scenario, leading to severe constraints on the $f(T)$
gravity models as an explanation to the cosmic acceleration. 
\end{abstract}


\maketitle

\section{Introduction}

\label{sect:Introduction}

There has been much recent interest in the so-called $f(T)$ gravity theory
as an alternative to dark energy for explaining the acceleration of the
universe \cite{Bengochea:2009, Yu:2010a, Myrzakulov:2010a, Tsyba:2010,
Linder:2010, Yi:2010b, Kazuharu:2010a, Kazuharu:2010b, Myrzakulov:2010b,
Yu:2010c, Karami:2010, Dent:2010, Li:2010, Zheng:2010, Dent:2011,
Sotiriou:2010, Zhang:2011}. This theory is a generalisation of the
teleparallel gravity \cite{Einstein, Aldrovandi} created by replacing $T$,
the lagrangian  of teleparallel gravity, by a function  $f(T$). It uses the
curvature-free Weitzenbock connection \cite{Weitzenbock} to define the
covariant derivative instead of the conventional torsionless Levi-Civita
connection of general relativity

Teleparallel gravity (see Ref.~\cite{Aldrovandi} for a review and references
therein) has a set of four tetrad (or vierbein) fields which form the
orthogonal bases for the tangent space at each point of spacetime. They are
the dynamical variables and play the role of the metric tensor field in
general relativity. The vierbeins are parallel vector fields, which gave the
theory the descriptor ``teleparallel". It is dynamically equivalent to
general relativity and so is not really an alternative to it, but a
reformulation which allows for a different interpretation: gravity is not
due to curvature, but to torsion.

The generalisation to $f(T)$ gravity can be viewed as an phenomenological
extension of teleparallel gravity (which is the special case $f(T)=T$),
inspired by the $f(R)$ generalization (see Ref.~\cite{Sotiriou:2008rp} for a
review) of general relativity. However, it has the advantage over $f(R)$
gravity that its field equations are second-order instead of fourth-order
(although it is know that, even though it leads to fourth-order equations, $%
f(R)$ gravity can be ghost free). Yet, it also possesses disadvantages.
Although $f(R)$ gravity is probably not the low-energy limit of some
fundamental theory, it does include models that can be motivated by
effective field theory. In contrast, $f(T)$ gravity seems at this stage to
be just an \emph{ad hoc} generalization.

Another serious disadvantage of $f(T)$ gravity that was pointed out very
recently in Ref.~\cite{Li:2010,Sotiriou:2010} is that it does not respect
local Lorentz symmetry. From a theoretical perspective this is a rather 
undesirable feature and experimentally there are stringent constraints. A
Lorentz-violating theory is only attractive if the violations are small
enough to avoid detection andit leads to some other significant
pay-off. So far, the only such pay-off that has been suggested is that $f(T)$
gravity might provide an alternative to conventional dark energy in general
relativistic cosmology. 

The specific models that have been considered in the literature \cite%
{Bengochea:2009, Yu:2010a, Myrzakulov:2010a, Tsyba:2010, Linder:2010,
Yi:2010b, Kazuharu:2010a, Kazuharu:2010b, Myrzakulov:2010b, Yu:2010c,
Karami:2010, Dent:2010} are rather special as they are tailored to
reproduce the late-time accelerated expansion of the universe without a
cosmological constant. However, to do so, a parameter in these models is
required to be tuned to a very low value, comparable with the observed value
of the cosmological constant. Thus, given the lack of clear theoretical
motivation for these models, it is rather questionable if this can really be
considered to be a resolution of the cosmological constant problem.

Nonetheless, given the attention these models have attracted recently, it
seems worthwhile to address their viability as alternatives to general
relativity in the field of cosmology itself, which was their initial
motivation. We do so by going beyond background cosmology and considering
linear perturbations. We will show that these models behave very differently
from the $\Lambda $CDM model on large scales, and are, therefore, very
unlikely to be suitable alternatives to it.

Cosmological perturbations in these models have  been considered recently in 
Refs.~\cite{Dent:2010, Zheng:2010, Dent:2011} as well. However, in these papers the field equations are written with only
partial derivatives and look quite different from the usual Einstein
equations. Here, we will present a covariant version of the field equations of  $f(T)$ gravity with a clear correspondence to the Einstein's equations. We will use them to derive the field
equations for a perturbed Friedman--Lema\^itre--Robertson--Walker  (FLRW) universe. As $f(T)$ gravity has different
dynamical variables (the tetrad fields) from general relativity
or $f(R)$ gravity (rank-2 tensorial metric field), we end up with
new degree(s) of freedom. This fact has been neglected before, for example, 
in Refs.~\cite{Dent:2010, Dent:2011}. Here we will show how the new degree of freedom
arise in the perturbed field equations, and numerically assess its effects on the linear-perturbation
observables, such as the CMB and matter power spectra.


The paper is arranged as follows: in Sect.~\ref{sect:Equations} we derive
the field equations for $f(T)$ gravity in its original form and rewrite
it in the covariant form. In Sect.~\ref{sect:pert} we give a detailed
introduction to the method of deriving the covariant and gauge-invariant
linear perturbation equations for $f(T)$ gravity theory, and list those
equations. We focus on a specific model with a power-law functional form for 
$f(T)$ in Sect.~\ref{sect:num}, and study its background cosmology and the
growth of large-scale structure. We summarise and conclude in Sect.~\ref%
{sect:con}. Throughout this work we use the metric convention $(+,-,-,-)$
and set $c=1$ and $\kappa =8\pi G$, where $G$ is the gravitational constant.

\section{The $f(T)$ Model and its Equations}

\label{sect:Equations}

In this section we give a brief description of the $f(T)$ model and a
detailed derivation of its field equations. In contrast to previous works,
which wrote these equations in terms of partial derivatives of the tetrads, we
shall do this by expressing them in terms of the Einstein tensor plus
relevant covariant derivatives of the vierbein field. This approach makes
the equations closely resemble their counterparts in GR and provides a basis
for the derivation of the perturbation equations in the  and gauge covariant
formalism, which is our final goal.

Since $f(T)$ gravity is a simple generalisation of teleparallel gravity
theory, we shall briefly introduce the latter (for a comprehensive review
see \cite{Aldrovandi}).
\subsection{Ingredients of Teleparallel Gravity}

\label{subsect:model}

In teleparallel gravity we have the vierbein, or tetrad, fields, $\mathbf{h}%
_{a}\left( x^{\mu }\right) $, as our dynamical variables; Latin indices $%
a,b,\cdots $ run from $0$ to $3$ and label tangent-space coordinates; Greek
indices $\mu ,\nu ,\cdots $ run from $0$ to $3$ and label spacetime
coordinates. The $h_{a}^{\mu }$ are both spacetime vectors and Lorentz
vectors in the tangent space. As the former (indexed by $\mu $), they are
the dynamical fields of gravitation, as the latter (indexed by $a$), they
form an orthonormal basis for the tangent space at each spacetime point.

The metric tensor of spacetime, $g_{\mu\nu}$, is given by
\begin{eqnarray}\label{eq:metric}
g_{\mu\nu} = \eta_{ab}h^a_\mu h^b_\nu
\end{eqnarray}
where $\eta_{ab}={\rm diag}(1,-1,-1,-1)$ is the Minkowski metric
for the tangent space. From this relation it follows that
\begin{eqnarray}
h_a^\mu h_\nu^a\ =\ \delta^\mu_\nu, \ \ \ h_a^\mu h^b_\mu\ =\ \delta^b_a,
\end{eqnarray}
where Einstein convention of summation has been used.
Eq.~(\ref{eq:metric}) implies that in this model
$g_{\mu\nu}$, $h_a^\mu$ and $h_\mu^a$ are all dependent on each
other, which is important for the derivation of
the field equations by variation.

General relativity  is built on the
Levi-Civita connection of the metric 
\begin{eqnarray}\label{eq:levi-civita}
\Gamma^\alpha_{\beta\gamma} &\equiv& \frac{1}{2}g^{\alpha\lambda}\left(g_{\lambda\beta,\gamma} + g_{\lambda\gamma,\beta} - g_{\beta\gamma,\lambda}\right),
\end{eqnarray}
where a comma is used to denote a partial derivative
($_{,\mu}\equiv\partial/\partial x^\mu$). This connection has
nonzero curvature but zero torsion. Teleparallel
gravity, or the teleparallel interpretation of general relativity, instead makes use of the Weitzenbock connection (tilded to
distinguish from $\Gamma^\alpha_{\beta\gamma}$)
\begin{eqnarray}\label{eq:weizenbock}
\tilde{\Gamma}^\alpha_{\ \beta\gamma} &\equiv& h^\alpha_b\partial_\gamma h^b_\beta\ =\ -h^b_\beta\partial_\gamma h^\alpha_b
\end{eqnarray}
which has a zero curvature but nonzero torsion. The
torsion tensor reads
\begin{eqnarray}\label{eq:torsion}
T^\alpha_{\ \beta\gamma} &\equiv& \tilde{\Gamma}^\alpha_{\ \gamma\beta}  - \tilde{\Gamma}^\alpha_{\ \beta\gamma}\
=\ h_b^\alpha\left(\partial_\beta h_\gamma^b - \partial_\gamma h^b_\beta\right).
\end{eqnarray}
The difference between the Levi-Civita and Weitzenbock
connections, neither of which is a spacetime tensor, is a spacetime tensor, and is
known as the contorsion tensor:
\begin{eqnarray}\label{eq:contorsion}
K^\rho_{\mu\nu} &\equiv& \tilde{\Gamma}^\rho_{\ \mu\nu} - \Gamma^\rho_{\mu\nu}\
=\ \frac{1}{2}\left(T_{\mu\ \nu}^{\ \rho} + T_{\nu\ \mu}^{\ \rho} - T^{\rho}_{\ \mu\nu}\right).
\end{eqnarray}
It is worth pointing out at this point that, based on the definition we have given above, the Weitzenbock connection, the torsion tensor and the contorsion tensor are not local Lorentz scalars ({\em i.e.}~they do not remain invariant under a local Lorentz transformation in tangent space) even though they do not have any tangent space indices.\footnote{Teleparallelism assumes the existence of a class of frames where the spin connection is zero and in which the Weitzenbock connection assume the form given in Eq.~(\ref{eq:weizenbock}). We choose to work in one of these frame. One could introduce a Lorentz covariant formulation of the theory at the level of the action, but this would only change appearances \cite{Sotiriou:2010}.} This is the root of the lack of Lorentz invariance in generalized teleparallel theories of gravity.

The Lagrangian density of teleparallel gravity is given by
\begin{eqnarray}\label{eq:T}
\mathcal{L}_T &\equiv& \frac{h}{16\pi G}T\,,
\end{eqnarray}
where
\begin{eqnarray}
T = \frac{1}{4}T^{\rho\mu\nu}T_{\rho\mu\nu} + \frac{1}{2}T^{\rho\mu\nu}T_{\nu\mu\rho} - T_{\rho\mu}^{\ \ \rho}T^{\nu\mu}_{\ \ \ \nu},
\end{eqnarray}
and $h\equiv\sqrt{-g}$ is the determinant of $h_a^{\alpha}$
with $g$ being the determinant of the metric $g_{\mu\nu}$. After adding the matted Lagrangian density $\mathcal{L}_m$, variation with respect to the tetrad yields the field equations
\begin{eqnarray}\label{eq:field_eqn_tel}
\partial_\rho \left(hh_a^\nu S_{\nu}^{\ \lambda\rho}\right)&-&hh^\rho_a S^{\mu\nu\lambda}T_{\mu\nu\rho} \nonumber\\
&+&  \frac{1}{4}hh^\lambda_a S^{\rho\mu\nu}T_{\rho\mu\nu}=8\pi G\Theta_a^{\lambda}
\end{eqnarray}
with $\Theta_a^{\lambda}$ related to the usual energy-momentum tensor
$\Theta_{\mu\nu}$ by
$\Theta^{\mu\nu}\equiv\eta^{ab}\Theta_{a}^{\nu}h_b^\mu$ and
\begin{eqnarray}\label{eq:S}
S^{\rho\mu\nu} &\equiv& K^{\mu\nu\rho} - g^{\rho\nu}T^{\sigma\mu}_{\ \ \ \sigma} + g^{\rho\mu}T^{\sigma\nu}_{\ \ \ \sigma}.
\end{eqnarray}

\subsection{Field Equation for $f(T)$ Gravity}

\label{subsect:f(T)}

The idea of $f(T)$ gravity is simply to promote the $T$ in the
Lagrangian to become an arbitrary function of $T$:
\begin{eqnarray}
\mathcal{L}_T &\rightarrow& \mathcal{L}\ =\ \frac{h}{16\pi G}f(T).
\end{eqnarray}
The field equations are straightforward generalizations of those of standard teleparallel gravity just given above:
\begin{eqnarray}\label{eq:field_eqn}
f_T\left[\partial_\rho \left(hh_a^\nu S_{\nu}^{\
\lambda\rho}\right) - hh^\rho_a
S^{\mu\nu\lambda}T_{\mu\nu\rho}\right] \nonumber\\ +
f_{TT}hh_a^\nu S_{\nu}^{\ \lambda\rho}\partial_\rho T +
\frac{1}{2}hh^\lambda_a f(T) &=& 8\pi G\Theta_a^{\lambda}
\end{eqnarray}
where $f_T\equiv\partial f(T)/\partial T$ and
$f_{TT}\equiv\partial^2f(T)/\partial T^2$.

Obviously, if $f(T)=T+\Lambda$ with $\Lambda$ a constant, then
Eq.~(\ref{eq:field_eqn}) simply reduces to
Eq.~(\ref{eq:field_eqn_tel}).

\subsection{Covariant Version of the Field Equations}

\label{subsect:gr-form}

The field equation Eqs.~(\ref{eq:field_eqn_tel}) and (\ref{eq:field_eqn}) are
written in terms of the tetrad and partial derivatives and appear very different
from Einstein's equations. This makes comparison with general
relativity rather difficult. In this subsection we show that Eq.~(\ref{eq:field_eqn_tel}) can be written in terms of the metric only and it then becomes Einstein's equation. We also present an equation relating $T$ with the Ricci scalar of the metric $R$. These will make the equivalence between teleparallel gravity and general relativity clear.
On the other hand, the tetrad cannot be eliminated completely in favour of the metric in Eq.~(\ref{eq:field_eqn}), because of the lack of local Lorentz symmetry, but we will show that the later can be brought in a form that closely resembles
Einstein's equation. This form is more suitable for introducing the covariant and gauge invariant formalism for cosmological perturbations.

First, let us note that although in
Sect.~\ref{subsect:model} the tensors were all written in terms of
partial derivatives, they could be rearranged so that all the
partial derivatives are replaced with covariant derivatives
compatible with the metric $g_{\mu\nu}$, {\it i.e.}, $\nabla_\alpha$
where $\nabla_\alpha g_{\mu\nu}=0$. In particular, we would have
\begin{eqnarray}
T^{\alpha}_{\ \beta\gamma} &=& h_b^\alpha\left(\partial_\beta h_\gamma^b - \Gamma^\sigma_{\beta\gamma}h^b_\sigma - \partial_\gamma h^b_\beta + \Gamma^\sigma_{\gamma\beta}h^b_\sigma\right)\nonumber\\
&=& h_b^\alpha\left(\nabla_\beta h_\gamma^b - \nabla_\gamma
h^b_\beta\right),
\end{eqnarray}
where we have used the fact that $\Gamma^\alpha_{\beta\gamma}$ is
symmetric in the subscripts $\beta, \gamma$. From this it can be
readily checked that
\begin{eqnarray}\label{eq:K_cov}
K^\beta_{\ \gamma\alpha} &=& h^\beta_a\nabla_\alpha h^a_\gamma,\\
\label{eq:S_cov}
S_\alpha^{\ \beta\gamma} &=& \eta^{ab}h^\beta_a\nabla_\alpha h_b^\gamma + \delta^\gamma_\alpha\eta^{ab}h^\mu_a\nabla_\mu h_b^\beta\nonumber\\&&\qquad - \delta^\beta_\alpha\eta^{ab}h^\mu_a\nabla_\mu h^\gamma_b
\end{eqnarray}
and clearly
\begin{eqnarray}
K^{\alpha\beta\gamma} &=& K^{[\alpha\beta]\gamma},\nonumber\\
T^{\alpha\beta\gamma} &=& T^{\alpha[\beta\gamma]},\nonumber\\
S^{\alpha\beta\gamma} &=& S^{\alpha[\beta\gamma]},\nonumber
\end{eqnarray}
in which the square brackets mean anti-symmetrisation, and also
\begin{eqnarray}
K^{\mu}_{\ \rho\mu}\ =\ T^{\mu}_{\ \mu\rho}, \ \ \ K^{\rho\mu}_{\
\ \ \mu}\ =\ T^{\mu\rho}_{\ \ \ \mu}.\nonumber
\end{eqnarray}
These relations are useful in the calculation below.

Next, from the relation between $\Gamma^\alpha_{\beta\gamma}$ and
$\tilde{\Gamma}^\alpha_{\beta\gamma}$ as given in
Eq.~(\ref{eq:contorsion}) and the fact that the
curvature tensor associated with the Weitzenbock connection
$\tilde{\Gamma}^\alpha_{\beta\gamma}$ vanishes, we can
write the Riemann tensor for the connection
$\Gamma^\alpha_{\beta\gamma}$ as \cite{Aldrovandi}
\begin{eqnarray}
R^\rho_{\ \mu\lambda\nu} &=& \partial_\lambda\Gamma^\rho_{\mu\nu} - \partial_\nu\Gamma^{\rho}_{\mu\lambda} + \Gamma^{\rho}_{\sigma\lambda}\Gamma^\sigma_{\mu\nu} - \Gamma^\rho_{\sigma\nu}\Gamma^\sigma_{\mu\lambda}\nonumber\\
&=& \nabla_\nu K^\rho_{\ \mu\lambda} - \nabla_\lambda K^\rho_{\
\mu\nu} + K^\rho_{\ \sigma\nu}K^\sigma_{\ \mu\lambda} -
K^{\rho}_{\ \sigma\lambda}K^\sigma_{\ \mu\nu}\nonumber
\end{eqnarray}
and the corresponding Ricci tensor and Ricci scalar are
\begin{eqnarray}\label{eq:Ricci}
R^\rho_{\ \lambda}
&=& \nabla_\mu K^{\rho\mu}_{\ \ \ \lambda} - \nabla_\lambda K^{\rho\mu}_{\ \ \ \mu} + K^\rho_{\ \sigma\mu}K^{\sigma\mu}_{\ \ \ \lambda} - K^{\rho}_{\ \sigma\lambda}K^{\sigma\mu}_{\ \ \ \mu}\nonumber\\
\label{eq:Ricci2}R &=& K^{\mu\rho}_{\ \ \ \mu}K^\nu_{\ \rho\nu} - K^{\mu\nu\rho}K_{\rho\nu\mu} - 2\nabla^\mu\left(T^\nu_{\ \mu\nu}\right)\nonumber\\
&=& -T - 2\nabla^\mu\left(T^\nu_{\ \mu\nu}\right).
\end{eqnarray}
This last equation implies that the $T$ and $R$ differ only by a covariant divergence of a spacetime vector. Therefore, the Einstein-Hilbert action and the teleparallel action ({\em i.e.}~the action constucted with the Lagrangian density given in Eq.~(\ref{eq:T})) will both lead to the same field equations and are dynamically equivalent theories. 

We can show this equivalence directly at the level of the field equations. With the aid of the equations listed above, it can be shown, after some
algebraic manipulations, that 
\begin{equation}\label{eq:einstein-tensor}
hh_a^\rho G^\lambda_{\ \rho} 
= \partial_\xi\left(hh_a^\rho S_\rho^{\ \lambda\xi}\right) - hh_a^\xi S^{\mu\nu\lambda}T_{\mu\nu\xi}+ \frac{1}{2}hh_a^\lambda T,
\end{equation}
where $G_{\mu\nu}$ is the Einstein tensor.
Substituting this equation into Eq.~(\ref{eq:field_eqn_tel}) and rearranging, we
obtain Einstein's equations. If we do the same for 
Eq.~(\ref{eq:field_eqn}) we get
\begin{eqnarray}\label{eq:modified_einstein_eqn}
f_TG_{\mu\nu} &+& \frac{1}{2}g_{\mu\nu}\left[f(T)-f_TT\right] \nonumber\\&&\qquad+
f_{TT}S_{\nu\mu\rho}\nabla^\rho T = 8\pi G\Theta_{\mu\nu}.
\end{eqnarray}
Eq.~(\ref{eq:modified_einstein_eqn}) can be taken as the
starting point of the $f(T)$ modified gravity model, and it has a
structure similar to the field equation of the $f(R)$ gravity.
Note that when $f(T)=T$ general relativity is exactly recovered, as expected.

\section{The Linear Perturbation Equations}

\label{sect:pert}

In order to study the evolution of linear perturbations in the $f(T)$
gravity, we have to linearise the field equations. Usually, this is achieved
by writing all quantities in terms of the metric perturbation variables, and
for this we have to use some metric ansatz. In $f(T)$ gravity, however, it
is the vierbein, rather than the metric, that is the fundamental field, and
it has 16 rather than 10 independent components. Usually, the six additional
components correspond to local Lorentz symmetry, but, as mentioned
previously $f(T)$ gravity is not invariant under that symmetry.
Consequently, specifying a metric ansatz does not necessarily fix all the
tetrad components \cite{Li:2010}, and one needs to specify an ansatz for the
tetrad itself and derive the metric perturbations thereafter.

Other subtleties emerge here. For example, one cannot write the metric in
some familiar gauges ({\em e.g.}~conformal Newtonian) \textit{a priori}, as these
gauges are obtained by gauging away certain degrees of freedom in the metric
fields. However, in $f(T)$ gravity the degrees of freedom are different and
the lack of local Lorentz invariance means that we can only gauge away 4 of
the 16 components of the tetrad due to the invariance of the action under
spacetime coordinate transformations. 

We follow Ref.~\cite{GR3+1} and derive the perturbation equations in the $3+1
$ formalism, in which all the quantities are covariant and gauge invariant.
This formalism deals with physical quantities directly and does not need to
make a metric ansatz \textit{a priori}. It is appropriate to use it given
that we have derived the field equation Eq.~(\ref{eq:modified_einstein_eqn})
in the covariant form. It has proved quite useful in studies of perturbation
evolution in modified gravity theories \cite{Li:pfr-1, Li:pfr-2, Li:mfr,
Li:mgb, Li:aether}.

\subsection{Covariant and Gauge Invariant Perturbation Equations in General Relativity}

\label{subsect:GR3+1}

The $3+1$ decomposition makes spacetime splits of physical quantities with respect to the 4-velocity $u^{\alpha}$ of an observer. The projection tensor $H_{\alpha\beta}$ is defined as $H_{\alpha\beta}=g_{\alpha\beta}-u_{\alpha}u_{\beta}$ and can be used to obtain covariant tensors perpendicular to $u$. For example, the covariant spatial derivative $\hat{\nabla}$ of a tensor field $T_{\mu\cdot\cdot\cdot\nu}^{\beta\cdot\cdot\cdot\gamma}$ is defined as
\begin{eqnarray}\label{eq:spatial_deriv}
\hat{\nabla}^{\alpha}T_{\mu\cdot \cdot \cdot\nu}^{\beta\cdot \cdot \cdot \gamma}\equiv
H_{\rho}^{\alpha}H_{\sigma}^{\beta}\cdot \cdot \cdot \ H_{\delta}^{\gamma}H_{\mu}^{\lambda}\cdot \cdot \cdot \
H_{\nu}^{\xi}\nabla ^{\rho}T_{\lambda\cdot \cdot \cdot \xi}^{\sigma\cdot \cdot \cdot \delta}.
\end{eqnarray}
The energy-momentum tensor and covariant derivative of the 4-velocity are
decomposed respectively as
\begin{eqnarray}\label{eq:decomp1}
\Theta_{\alpha\beta} &=& \pi_{\alpha\beta}+2q_{(\alpha}u_{\beta)}+\rho u_{\alpha}u_{\beta}-pH_{\alpha\beta},\\
\label{eq:decomp2}\nabla _{\alpha}u_{\beta} &=&
\sigma_{\alpha\beta}+\varpi_{\alpha\beta}+\frac{1}{3}\theta
H_{\alpha\beta}+u_{\alpha}A_{\beta}.
\end{eqnarray}
In the above expressions, $\pi_{\alpha\beta}$ is the projected symmetric trace-free (PSTF)
anisotropic stress, $q_{\alpha}$ the heat flux vector, $p$ the isotropic
pressure, $\sigma _{\alpha\beta}$ the PSTF shear tensor, $\varpi _{\alpha\beta}=\hat{\nabla}_{[\alpha}u_{\beta]}$ the vorticity, $\theta = \nabla^{\alpha}u_{\alpha} = 3\dot{a}/a$ ($a$ is the mean expansion scale factor) the expansion scalar, and $A_{\alpha}=\dot{u}_{\alpha}$ the acceleration; the overdot denotes time derivative expressed as $\dot{\phi}=u^{\alpha}\nabla_{\alpha}\phi $, brackets mean antisymmetrisation, and parentheses symmetrization. The 4-velocity normalization is chosen to be $ u^{\alpha}u_{\alpha}=1$. The quantities $\pi_{\alpha\beta}, q_{\alpha}, \rho, p$ are referred to as \emph{dynamical} quantities and $\sigma_{\alpha\beta}, \varpi_{\alpha\beta}, \theta, A_{\alpha}$ as \emph{kinematical} quantities. Note that the dynamical quantities can also be obtained from the energy-momentum tensor $\Theta_{\alpha\beta}$ through the relations
\begin{eqnarray}\label{eq:dynamic_quantities}
\rho &=& \Theta_{\alpha\beta}u^{\alpha}u^{\beta},\nonumber\\
p &=& -\frac{1}{3}H^{\alpha\beta}\Theta_{\alpha\beta},\nonumber\\
q_{\alpha} &=& H_{\alpha}^{\mu}u^{\mu}\Theta_{\mu\nu},  \nonumber \\
\pi_{\alpha\beta} &=& H_{\alpha}^{\mu}H_{\beta}^{\nu}\Theta_{\mu\nu}+pH_{\alpha\beta}.
\end{eqnarray}

Decomposing the Riemann tensor and making use the Einstein equations,  after linearisation we
obtain five constraint equations \cite{GR3+1}:
\begin{eqnarray}
\label{eq:constraint_varpi} 0 &=& \hat{\nabla}^{\mu}(\varepsilon _{\ \ \mu\nu}^{\alpha\beta}u^{\nu}\varpi_{\alpha\beta});\\
\label{eq:constraint_q} \kappa q_{\alpha} &=& -\frac{2\hat{\nabla}_{\alpha}\theta}{3} + \hat{\nabla}^{\beta}\sigma_{\alpha\beta}+\hat{\nabla}^{\beta}\varpi_{\alpha\beta};\ \ \  \\
\mathcal{B}_{\alpha\beta} &=& \left[ \hat{\nabla}^{\mu}\sigma_{\nu(\alpha}+\hat{\nabla}^{\mu}\varpi_{\nu(\alpha}\right] \varepsilon_{\beta)\rho\mu}^{\ \ \ \ \nu}u^{\rho};\\
\label{eq:constraint_phi} \hat{\nabla}^{\beta}\mathcal{E}_{\alpha\beta} &=& \frac{1}{2}\kappa \left[\hat{\nabla}^{\beta}\pi_{\alpha\beta}+\frac{2}{3}\theta q_{\alpha}+\frac{2}{3}\hat{\nabla}_{\alpha}\rho\right];\\
\hat{\nabla}^{\beta}\mathcal{B}_{\alpha\beta} &=& \frac{1}{2}\kappa\left[\hat{\nabla}_{\mu}q_{\nu}+(\rho+p)\varpi_{\mu\nu}\right]\varepsilon_{\alpha\beta}^{\ \ \mu\nu}u^{\beta},
\end{eqnarray}
and five propagation equations:
\begin{eqnarray}
\label{eq:raychaudhrui} \dot{\theta}+\frac{1}{3}\theta^{2}
-\hat{\nabla}^{a}A_{a}+\frac{\kappa }{2}(\rho +3p) &=& 0; \\
\label{eq:propagation_sigma} \dot{\sigma}_{\alpha\beta}+\frac{2}{3}\theta\sigma_{\alpha\beta}-\hat{\nabla}_{\langle\alpha}A_{\beta\rangle }+\mathcal{E}_{\alpha\beta}+\frac{1}{2}\kappa\pi_{\alpha\beta} &=& 0; \\
\dot{\varpi}+\frac{2}{3}\theta \varpi -\hat{\nabla}_{[\alpha}A_{\beta]} &=& 0; \\
\label{eq:propagation_phi} \frac{1}{2}\kappa\left[\dot{\pi}_{\alpha\beta}+\frac{1}{3}\theta\pi_{\alpha\beta}\right] - \frac{1}{2}\kappa\left[(\rho+p)\sigma_{\alpha\beta}+\hat{\nabla}_{\langle\alpha}q_{\beta\rangle}\right] \nonumber\!\!\!\!\!\!\!\!\!\!\!\!\!\\
-\left[\dot{\mathcal{E}}_{\alpha\beta}+\theta\mathcal{E}_{\alpha\beta}-\hat{\nabla}^{\mu}\mathcal{B}_{\nu(\alpha}\varepsilon_{\beta)\rho\mu}^{\ \ \ \ \nu}u^{\rho}\right] &=& 0;\ \ \ \ \\
\dot{\mathcal{B}}_{\alpha\beta}+\theta\mathcal{B}_{\alpha\beta}+\hat{\nabla}^{\mu}\mathcal{E}_{\nu(\alpha}\varepsilon_{\beta)\rho\mu}^{\ \ \ \ \nu}u^{\rho}\nonumber\\+\frac{\kappa}{2}\hat{\nabla}^{\mu}\mathcal{\pi}_{\nu(\alpha}\varepsilon_{\beta)\rho\mu}^{\ \ \ \ \nu}u^{\rho} &=& 0.
\end{eqnarray}
Here, $\varepsilon_{\alpha\beta\mu\nu}$ is the covariant permutation tensor, $\mathcal{E}_{\alpha\beta}$ and $\mathcal{B}_{\alpha\beta}$ are respectively the electric and magnetic parts of the Weyl tensor $\mathcal{W}_{\alpha\beta\mu\nu}$, defined by $\mathcal{E}_{\alpha\beta}=u^{\mu}u^{\nu}\mathcal{W}_{\alpha\mu\beta\nu}$ and $\mathcal{B}_{\alpha\beta}=-\frac{1}{2}u^{\mu}u^{\nu}\varepsilon_{\alpha\mu}^{\ \ \rho\sigma}\mathcal{W}_{\rho\sigma\beta\nu}$. Note that the angle bracket denotes taking the trace-free part of a quantity.

Using the definition of the projected derivatives, it can be shown that
\begin{eqnarray}
[\hat{\nabla}_\mu,\hat{\nabla}_\nu]v_\rho &=& - 2\varpi_{\mu\nu}\dot{v}_\rho +H^\alpha_\mu H^\beta_\nu H^\gamma_\rho R_{\alpha\beta\gamma}^{\ \ \ \ \lambda}v_\lambda\nonumber\\ 
&&+\left(\hat{\nabla}_\mu u_\rho\hat{\nabla}_\nu u^\lambda-\hat{\nabla}_\mu u^\lambda\hat{\nabla}_\nu u_\rho\right)v_\lambda 
\end{eqnarray}
for any projected vector field $v_\rho$ ($u^\rho v_\rho = 0$). In the absence of vorticity $\varpi_{\mu\nu}$ (which is true up to first order in perturbation because we are considering the scalar mode only), the above equation can be written as
\begin{eqnarray}
[\hat{\nabla}_\mu,\hat{\nabla}_\nu]v_\rho &\equiv& -\hat{R}_{\mu\nu\rho}^{\ \ \ \ \lambda}v_\lambda
\end{eqnarray}
where $\hat{R}_{\mu\nu\rho\lambda}$ is the spatial 3-curvature tensor defined in analogy to the Riemann curvature tensor in the 4D spacetime (the minus sign is conventional). We can then define the corresponding Ricci scalar of the hyperspace perpendicular to the 4-velocity in the usual way: $\hat{R}=\hat{R}_{\mu\nu}^{\ \ \mu\nu}$. With the Einstein equation it is easy to find
\begin{eqnarray}\label{eq:spatial_ricci}
\hat{R} &\approx& 2\kappa\rho - \frac{2}{3}\theta ^{2}.
\end{eqnarray}
The spatial derivative of $\hat{R}$, $\eta _{\alpha} \equiv \frac{1}{2}a\hat{\nabla}_{\alpha}\hat{R}$, is then given as
\begin{eqnarray}\label{eq:constraint_eta}
\eta _{\alpha} &=& \kappa\hat{\nabla}_{\alpha}\rho - \frac{2a}{3}\theta\hat{\nabla}_{\alpha}\theta,
\end{eqnarray}
and its propagation equation by
\begin{eqnarray}\label{eq:propagation_eta}
\dot{\eta}_{\alpha}+\frac{2\theta}{3}\eta_{\alpha} &=& -\frac{2}{3}\theta a\hat{\nabla}_{\alpha}\hat{\nabla}\cdot A - a\kappa\hat{\nabla}_{\alpha}\hat{\nabla}\cdot q.
\end{eqnarray}

Finally, there are the conservation equations for the energy-momentum tensor:
\begin{eqnarray}  \label{eq:energy_conservation}
\dot{\rho}+(\rho +p)\theta +\hat{\nabla}^{\alpha}q_{\alpha} &=& 0,\\
\label{eq:heat_flux_evolution} \dot{q}_{\alpha}+\frac{4}{3}\theta q_{\alpha}+(\rho+p)A_{\alpha}-\hat{\nabla}_{\alpha}p+\hat{\nabla}^{\beta}\pi_{\alpha\beta} &=& 0.
\end{eqnarray}

As we are considering a spatially-flat universe,\footnote{See Ref.~\cite{Ferraro:2011us} for a discussion on hyperspherical and hyperbolic universes.} the spatial
curvature must vanish on large scales and so in the background $\hat{R}=0$. Thus,
from Eq.~(\ref{eq:spatial_ricci}), we obtain
\begin{eqnarray}
\frac{1}{3}\theta^{2} &=& \kappa\rho.
\end{eqnarray}
This is the Friedmann equation in general relativity, and the other
background equations can be obtained by taking the zero-order parts of Eqs.~(\ref{eq:raychaudhrui}, \ref{eq:energy_conservation}), yielding:
\begin{eqnarray}
\dot{\theta}+\frac{1}{3}\theta ^{2}+\frac{\kappa }{2}(\rho +3p) &=& 0, \\
\label{eq:background_energy_conservation} \dot{\rho}+(\rho +p)\theta &=& 0.
\end{eqnarray}

\subsection{Generalisation to the $f(T)$ Gravity}

\label{subsect:fT3+1}

In order to make best use of the formulae obtained for general relativity,
we can consider the modifications to the Einstein equation in $f(T)$ gravity
as a new effective energy-momentum tensor $\Theta _{\mu \nu }^{eff}$ in
addition to that of the fluid matter, $\Theta _{\mu \nu }^{f}$.  Eq.~(\ref{eq:modified_einstein_eqn}) can then be rewritten as 
\begin{eqnarray}
G_{\mu\nu} &=& \kappa\left(\Theta^f_{\mu\nu} + \Theta^{eff}_{\mu\nu}\right)
\end{eqnarray}
in which
\begin{eqnarray}
\kappa\Theta^{eff}_{\mu\nu} &\equiv& -\frac{f_T-1}{f_T}\kappa\Theta^f_{\mu\nu} - \frac{1}{2f_T}g_{\mu\nu}\left[f-f_TT\right]\nonumber\\ &&- \frac{f_{TT}}{f_T}S_{\nu\mu\rho}\nabla^{\rho}T.
\end{eqnarray}
As already mentioned, here we have to work with the tetrad and not just the metric, so the setup will be slightly different than that of general relativity.

Since we intend to investigate the perturbation evolution in an almost
Friedmann universe, let us first consider an exact Friedmann universe: there
is no special spatial direction and the fundamental observer's world line is
in the time direction. Assuming that the comoving observer's frame is
aligned with the frame defined by the tetrad in tangent space we have $h_{%
\underline{0}}^{\mu }=u^{\mu }$, where $u^{\mu }$ is the 4-velocity of the
fundamental observer, and the $h_{\underline{i}}^{\mu }$ ($\underline{i}%
=1,2,3$) are three orthonormal vectors in the 3-space of the fundamental
observer (here we use an underline to denote components of the Lorentz
index). If we define $U_{a}\equiv h_{a}^{\mu }u_{\mu }$, then $1=g_{\mu \nu
}u^{\mu }u^{\nu }=\eta _{ab}h_{\mu }^{a}h_{\nu }^{b}u^{\mu }u^{\nu }=\eta
_{ab}U^{a}U^{b}$. Note that in this case $U^{a}=\delta _{\underline{0}}^{a}$.

In an almost Friedmann universe, the above symmetry is at best an
approximation, and $h_{\underline{0}}^{\mu }$ will not coincide exactly with 
$u^{\mu }$ but could differ slightly. Instead of $U^{a}=(1,\mathbf{0})$, we
will have $U^{a}=(U^{\underline{0}},U^{\underline{i}})$ where the $U^{%
\underline{i}}$ are small, and $\eta _{ab}U^{a}U^{b}=1$ implies that $U^{%
\underline{0}}=1$ up to first order in perturbation. As $U^{\underline{0}%
}=h_{\mu }^{\underline{0}}u^{\mu }$, we can write $h_{\mu }^{\underline{0}%
}=u_{\mu }+\ae _{\mu },$ where $\ae _{\mu }$ is a perturbation vector and $%
u^{\mu }\ae _{\mu }=0$. As it will turn out, all the information we need to
know about $h_{\mu }^{\underline{i}}$ is that $h_{\mu }^{\underline{i}%
}u^{\mu }=U^{\underline{i}}$ is first order in perturbation and $h_{\mu }^{%
\underline{i}}$. This suffices to show that, to this order of perturbation, $%
\ae _{\mu }$ is the only new physical degree of freedom with respect to
general relativity. It was expected to appear due to the lack of local
Lorentz invariance \cite{Li:2010}. Detailed calculations in support of these
statements, as well as explicit derivations of the perturbative expressions
for quantities entering the field equations can be found in the Appendix~\ref%
{appen:expression}. Here, we will only quote the results of these
calculations. Obviously, the only quantities that are not already present in
general relativity are $T$ and $S_{\nu \mu \rho }\nabla ^{\rho }T$. Up to
first order in perturbations, we have 
\begin{eqnarray}
\label{perT}
T &\approx& -\frac{2}{3}\theta^2-\frac{4}{3}\theta\hat{\nabla}\cdot\ae,\\
\label{perS}
S_{\nu\mu\rho}\nabla^{\rho}T &\approx& \frac{2}{3}\dot{T}\left(\theta+\hat{\nabla}\cdot\ae\right)H_{\mu\nu}-\frac{1}{2}\dot{T}u_{\nu}\hat{R}_{\mu}\nonumber\\
&&- \dot{T}\left(\sigma_{\mu\nu}+\varpi_{\mu\nu}+\hat{\nabla}_{\langle\mu}\ae_{\nu\rangle}+\hat{\nabla}_{[\mu}\ae_{\nu]}\right)\nonumber\\
&&-\frac{2}{3}\theta u_\mu\hat{\nabla}_{\nu}T.
\end{eqnarray}
Here $\hat{\nabla}\cdot\ae\equiv\hat{\nabla}^{\mu}\ae_{\mu}$ and $\hat{R}_{\mu}$ satisfies $\hat{\nabla}^\mu\hat{R}_\mu=\hat{R}$.

Using the definitions given in Eq.~(\ref{eq:dynamic_quantities}),
it is straightforward to obtain
\begin{eqnarray}
\label{eq:rho_eff}\kappa\rho^{eff} &\approx&
-\frac{1}{f_T}\left[(f_T-1)\kappa\rho^{f}+\frac{1}{2}\left(f-f_TT\right)\right],
\end{eqnarray}
\begin{eqnarray}
\label{eq:p_eff}\kappa p^{eff} &\approx&
-\frac{1}{f_T}\left[(f_T-1)\kappa p^{f}-\frac{1}{2}\left(f-f_TT\right)\right]\nonumber\\
&&+\frac{2}{3}\frac{1}{f_T}f_{TT}\dot{T}\left(\theta+\hat{\nabla}\cdot\ae\right),
\end{eqnarray}
\begin{eqnarray}
\label{eq:q_eff_1}\kappa q^{eff}_{\alpha} &\approx&
-\frac{1}{f_T}\left[(f_T-1)\kappa q^{f}_{\alpha} -
\frac{1}{2}f_{TT}\dot{T}\hat{R}_{\alpha}\right],\\
\label{eq:q_eff_2}&\approx& -\frac{1}{f_T}\left[(f_T-1)\kappa
q^{f}_{\alpha} -
\frac{2}{3}f_{TT}\theta\hat{\nabla}T\right],\\
\label{eq:pi_eff}\kappa\pi^{eff}_{\alpha\beta} &\approx&
-\frac{1}{f_T}\Big[(f_T-1)\kappa\pi^{f}_{\alpha\beta}-f_{TT}\dot{T}\left(\sigma_{\alpha\beta}
+\hat{\nabla}_{\langle\alpha}\ae_{\beta\rangle}\right)\Big].\nonumber\\
\end{eqnarray}
up to first order in perturbation. There are two different
expressions for $q^{eff}_{\alpha}$, which is because the quantity
$S_{\nu\mu\rho}\nabla^{\rho}T$ is not symmetric {\it a priori}, but the field equations require its antisymmetric part to vanish.

We are also interested in the density and pressure perturbations,
and these can be obtained by differentiating
Eqs.~(\ref{eq:rho_eff}, \ref{eq:p_eff}):
\begin{eqnarray}
\label{eq:delta_rho_eff}\kappa\hat{\nabla}_\alpha\rho^{eff}
&\approx&
\frac{1}{f_T}\left[(1-f_T)\kappa\hat{\nabla}_{\alpha}\rho^{f}+f_{TT}T\hat{\nabla}_{\alpha}T\right],\\
\label{eq:delta_p_eff}\kappa\hat{\nabla}_\alpha p^{eff} &\approx&
-\frac{1}{f_T}\Bigg[(f_T-1)\kappa\hat{\nabla}_{\alpha}p^{f}
 +\frac{8}{9}\theta^2\dot{\theta}f_{TTT}\hat{\nabla}_{\alpha}T\nonumber\\
&& -\frac{4}{3}\left(\dot{\theta}+\frac{2}{3}\theta^2\right)f_{TT}\hat{\nabla}_{\alpha}T
-\frac{2}{3}f_{TT}\theta\left(\hat{\nabla}_{\alpha}T\right)^{\cdot}
\nonumber\\&&+\frac{8}{9}\theta^2\dot{\theta}f_{TT}A_{\alpha}\Bigg].
\end{eqnarray}
Eqs.~(\ref{eq:q_eff_1}, \ref{eq:q_eff_2}, \ref{eq:pi_eff},
\ref{eq:delta_rho_eff}, \ref{eq:delta_p_eff}), together with the
equations given in Sect.~\ref{subsect:GR3+1}, are all we need to
study the perturbation evolution in $f(T)$ gravity.

\subsection{Scalar Equations in $f(T)$ Gravity}

\label{subsect:scalar_eqn}

Our formalism has so far been as general as possible. Now we will focus
exclusively on scalar perturbations and perform the following harmonic
expansions of our perturbation variables 
\begin{eqnarray}  \label{eq:HarmonicExpansion}
\hat{\nabla}_{\alpha}\rho = \sum_{k}\frac{k}{a}\mathcal{X}Q_{\alpha}^{k},\qquad \hat{%
\nabla}_{\alpha}p = \sum_{k}\frac{k}{a}\mathcal{X}^{p}Q_{\alpha}^{k}  \nonumber \\
q_{\alpha} = \sum_{k}qQ_{\alpha}^{k},\qquad \pi_{\alpha\beta} =
\sum_{k}\Pi Q_{\alpha\beta}^{k},\qquad
\nonumber \\
\hat{\nabla}_{\alpha}\theta = \sum_{k}\frac{k^{2}}{a^{2}}\mathcal{Z}Q_{\alpha}^{k}, \qquad \sigma_{\alpha\beta} = \sum_{k}\frac{k}{a}\sigma Q_{\alpha\beta}^{k}  \nonumber \\
\hat{\nabla}_{\alpha}a = \sum_{k}khQ_{\alpha}^{k},\qquad
A_{\alpha} = \sum_{k}\frac{k}{a}AQ^{k}_{\alpha}  \nonumber \\
\ae_{\alpha} = \sum_{k}\ae Q^{k}_{\alpha},\qquad \eta_{\alpha} = \sum_{k}\frac{k^{3}}{a^{2}%
}\eta Q_{\alpha}^{k}  \nonumber \\
\mathcal{E}_{\alpha\beta} = -\sum_{k}\frac{k^{2}}{a^{2}}\phi
Q_{\alpha\beta}^{k}
\end{eqnarray}
in which $Q^{k}$ is the eigenfunction of the comoving spatial Laplacian $%
a^{2}\hat{\nabla}^{2}$ satisfying
\begin{eqnarray}
\hat{\nabla}^{2}Q^{k} &=& \frac{k^{2}}{a^{2}}Q^{k}.\nonumber
\end{eqnarray}
$Q_{\alpha}^{k},Q_{\alpha\beta}^{k}$ are given by
$Q_{\alpha}^{k}=\frac{a}{k}\hat{\nabla}_{\alpha}Q^{k},
Q_{\alpha\beta}^{k}=\frac{a}{k}\hat{\nabla}_{\langle
\alpha}Q_{\beta\rangle }^{k}$.

In terms of the above harmonic expansion coefficients,
Eqs.~(\ref{eq:constraint_q}, \ref{eq:constraint_phi},
\ref{eq:propagation_sigma}, \ref{eq:propagation_phi},
\ref{eq:constraint_eta}, \ref{eq:propagation_eta}) can be
rewritten as \cite{GR3+1}
\begin{eqnarray}
\label{eq:constraint_q2} \frac{2}{3}k^{2}(\sigma - \mathcal{Z})
&=& \kappa qa^{2},\\
\label{eq:constraint_phi2} k^{3}\phi &=& -\frac{1}{2}\kappa
a^{2}\left[k(\Pi+\mathcal{X})+3\mathcal{H}q\right],\\
\label{eq:propagation_sigma2} k(\sigma' + \mathcal{H}\sigma) &=&
k^{2}(\phi+A)-\frac{1}{2}\kappa a^{2}\Pi,\\
\label{eq:propagation_phi2} k^{2}(\phi'+\mathcal{H}\phi) &=&
\frac{1}{2}\kappa a^{2}\left[k(\rho+p)\sigma+kq-\Pi'
-\mathcal{H}\Pi\right],\ \ \ \ \\
\label{eq:constraint_eta2} k^{2}\eta &=& \kappa\mathcal{X}a^{2} -
2k\mathcal{H}\mathcal{Z},\\
\label{eq:propagation_eta2} k\eta' &=& -\kappa qa^{2} -
2k\mathcal{H}A
\end{eqnarray}
in which $\mathcal{H}\equiv a'/a=\frac{1}{3}a\theta $ and a prime
denotes the derivative with respect to the conformal time $\tau$
($ad\tau =dt$). Also, Eq.~(\ref{eq:heat_flux_evolution}) and the
spatial derivative of Eq.~(\ref{eq:energy_conservation}) become
\begin{eqnarray}
\label{eq:heat_flux_evolution2} q' + 4\mathcal{H}q + (\rho+p)kA -
k\mathcal{X}^{p} +
\frac{2}{3}k\Pi &=& 0,\\
\label{eq:energy_conservation2} \mathcal{X}' + 3h'(\rho+p) +
3\mathcal{H}(\mathcal{X}+\mathcal{X}^{p}) + kq &=& 0
\end{eqnarray}
We shall always neglect the superscript
$^{\mathrm{tot}}$ for the total dynamical quantities and add
appropriate superscripts for individual matter species. Note that
\begin{eqnarray}
h' &=& \frac{1}{3}k\mathcal{Z} - \mathcal{H}A.
\end{eqnarray}
and $\rho, p, \mathcal{X}, \mathcal{X}^{p}, q, \Pi$ with
superscripts $^{f}$ or $^{eff}$ are the total quantities (fluid
matter plus correction terms). The harmonic coefficients
$\mathcal{X}^{eff}, \mathcal{X}^{p,eff}, q^{eff}, \Pi^{eff}$ can
be derived from Eqs.~(\ref{eq:delta_rho_eff},
\ref{eq:delta_p_eff}, \ref{eq:q_eff_1}, \ref{eq:q_eff_2},
\ref{eq:pi_eff}) such that

\begin{align}
\label{eq:delta_rho_tot}
  &f_T\kappa\left(\mathcal{X}^{f}+\mathcal{X}^{eff}\right)a^2
= \kappa\mathcal{X}^{f}a^2 +
24\frac{f_{TT}}{a^2}k\mathcal{H}^{3}(\mathcal{Z}+\ae),\\
\label{eq:delta_p_tot}
&f_T\kappa\left(\mathcal{X}^{p,f}+\mathcal{X}^{p,eff}\right)a^2
= \kappa\mathcal{X}^{p,f}a^2\nonumber\\&\qquad\qquad-\frac{f_{TT}}{a^2} [8k\mathcal{H}\left(3\mathcal{H}'-\mathcal{H}^2\right)(\mathcal{Z}+\ae)
\nonumber\\&\qquad\qquad+8k\mathcal{H}^2(\mathcal{Z}+\ae)' +24\mathcal{H}^2\left(\mathcal{H}'-\mathcal{H}^2\right)A]\nonumber\\&\qquad\qquad+96\frac{f_{TTT}}{a^4}k\mathcal{H}^3\left(\mathcal{H}'-\mathcal{H}^2\right)(\mathcal{Z}+\ae),
\\
\label{eq:q_tot_a}
& f_T\kappa\left(q^{f}+q^{eff}\right)a^2
= \kappa
q^{f}a^2 -8\frac{f_{TT}}{a^2}k^2\mathcal{H}^{2}(\mathcal{Z}+\ae)\\
\label{eq:q_tot_b}
& \qquad\qquad\qquad\qquad\;= \kappa q^{f}a^2 -
 12\frac{f_{TT}}{a^2}k\mathcal{H}\left(\mathcal{H}'-\mathcal{H}^2\right)\eta,\\
\label{eq:pi_tot}
& f_T\kappa\left(\Pi^{f}+\Pi^{eff}\right)a^2
= \kappa\Pi^{f}a^2\\&\qquad\qquad\qquad\qquad-12\frac{f_{TT}}{a^2}k\mathcal{H}\left(\mathcal{H}'-\mathcal{H}^2\right)(\sigma+\ae).\nonumber
\end{align}

This completes our derivation of the scalar mode covariant and
gauge-invariant perturbation equations for $f(T)$ gravity, and
we have one extra dynamical degree of freedom $\ae$. It is now
straightforward to choose a gauge, and as an example the
perturbation equations in the conformal Newtonian gauge are given
in Appendix~\ref{appen:newtonian}.

\begin{figure*}
\includegraphics[scale=1]{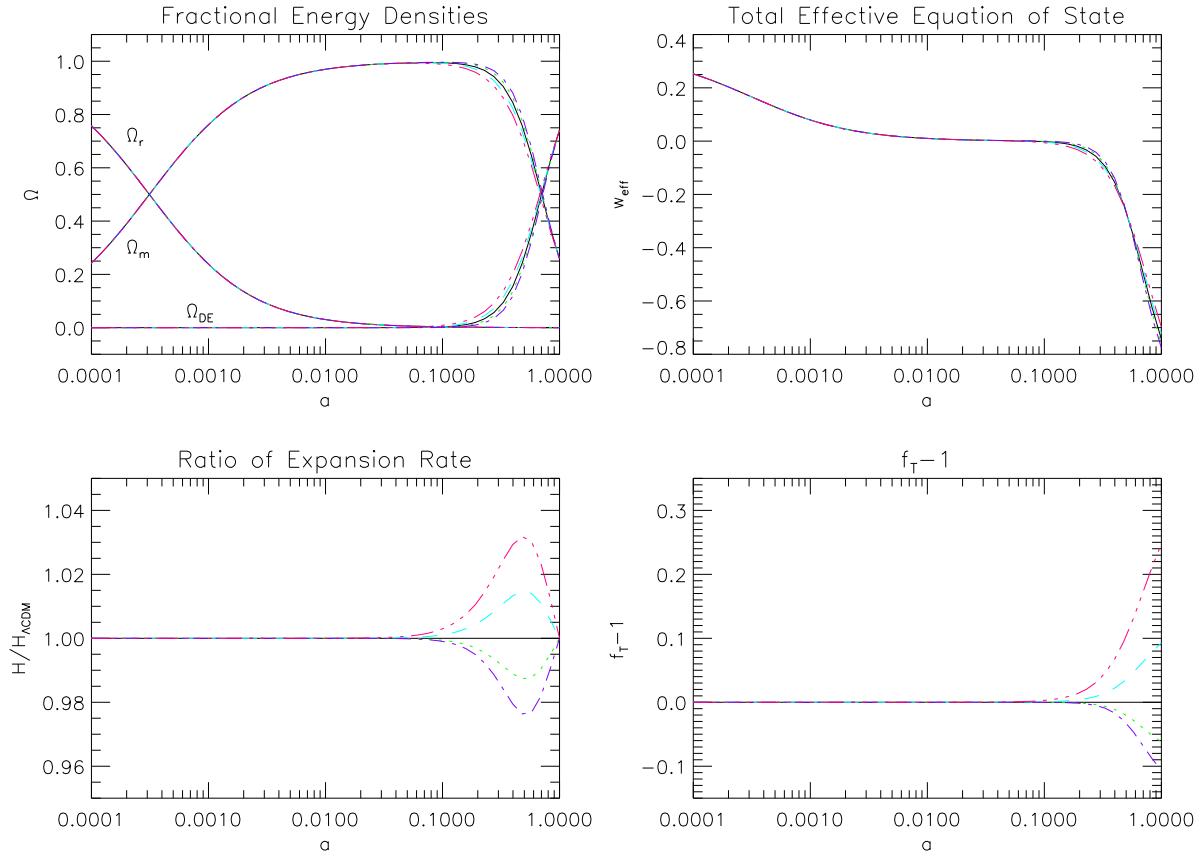}
\caption{(Colour online) The background evolution for the $f(T)$
gravity model with $f(T)=T-\mu^{2(1+n)}/(-T)^n$. {\it Upper-left
Panel}: the fractional energy densities for matter ($\Omega_m$),
radiation ($\Omega_r$) and the effective dark energy
($\Omega_{\mathrm{DE}}=1-\Omega_m-\Omega_r$), as functions of the
cosmic scale factor $a$, which is normalised to $1$ today. {\it
Upper-right Panel}: the total effective equation of state
$w_{eff}=-1-\frac{2\dot{H}}{3H^2}$, as a function of $a$. {\it
Lower-left Panel}: the ratio between the Hubble expansion rates
for the $f(T)$ gravity model and for the $\Lambda$CDM paradigm, as
a function of $a$. {\it Lower-right Panel}: $f_T-1$ as a function of
$a$. Here, results are shown for $n=0$ (black solid curve), $0.1$ (green dotted curve), $-0.1$ (cyan dashed curve), $0.2$ (purple dash-dotted curve) and $-0.2$ (pink dash-triple-dotted curve). Note that $n=0$ corresponds
to the $\Lambda$CDM paradigm. The relevant physical parameters are
$\Omega_m=0.257, \Omega_r=8.0331\times10^{-5}$ and
$H_0=71.9$~km/s/Mpc.} \label{fig:background}
\end{figure*}

\begin{figure}
\includegraphics[scale=0.49]{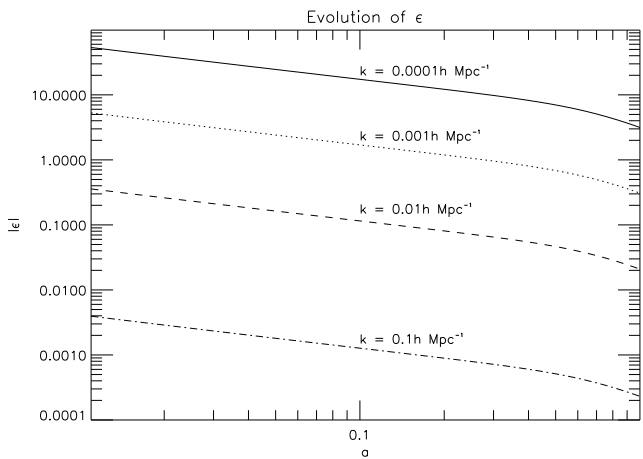}
\caption{The time-evolution of frame-independent quantity 
$\epsilon\equiv\ae+\sigma$, for the model with $f(T)=T-\mu^{2(1+n)}/(-T)^n$
and $n=0.1$, on different length scales, characterised respectively by
$k/(h~\mathrm{Mpc}^{-1})=10^{-4}$ (solid curve), $10^{-3}$ (dotted curve),
$10^{-2}$ (dashed curve) and $10^{-1}$ (dash-dotted curve).} \label{fig:ae}
\end{figure}

\begin{figure*}
\includegraphics[scale=1]{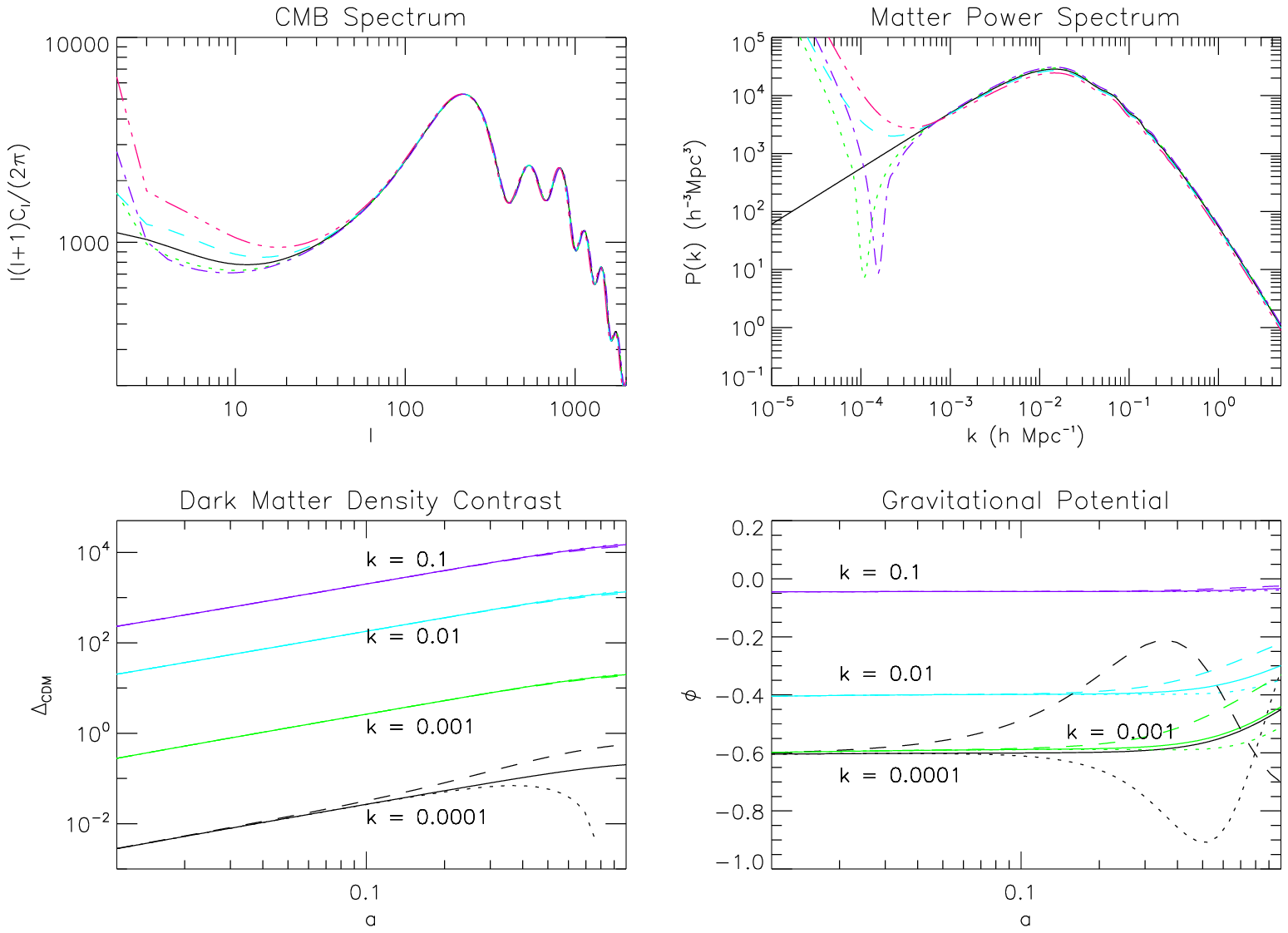}
\caption{(Colour online) The power spectra for the large-scale structure of the $f(T)$
gravity model with $f(T)=T-\mu^{2(1+n)}/(-T)^n$. {\it Upper-left
Panel}: the CMB spectrum for different values of $n$ -- $0$ (black
solid curve), $0.1$ (green dotted curve), $-0.1$ (cyan dashed
curve), $0.2$ (purple dash-dotted curve) and $-0.2$ (pink
dash-triple-dotted curve). {\it Upper-right Panel}: the same as
the upper-left panel, but for the matter power spectrum at
redshift $0$ (today). {\it Lower-left Panel}: the late-time
evolution of the dark matter density contrast
$\Delta_{\mathrm{CDM}}$ on different scales (as indicated besides
the curves); three values of $n$ have been considered -- $n=0.0$
(solid curves), $0.1$ (dotted curves) and $-0.1$ (dashed curves).
{\it Lower-right Panel}: the same as the lower-left panel, but for
the late-time evolution of the gravitational potential $\phi$ on
different scales. The physical parameters are the same as listed
in the caption of Fig.~\ref{fig:background}, and three species of
massless neutrinos are used.} \label{fig:linpert}
\end{figure*}

\section{Numerical Results}

\label{sect:num}

For a quantitative analysis of  the evolution of cosmological
perturbations in  $f(T)$ gravity, one needs to consider a concrete class of models. Since the motivation for considering $f(T)$ gravity was based on the suggestion that it could account for the late time cosmic speed-up without the need for dark energy, it makes sense to restricts ourselves to models that exhibit this property (we have expressed out reservations about the theoretical motivation of a general $f(T)$ theory in the Introduction). Thus, we focus on the class of models that can be parametrized as
\begin{eqnarray}
f(T) &=&T-\frac{\mu^{2(n+1)}}{(-T)^n}
\end{eqnarray}
where $n$ is some real number. The $\mu$ parameter will be fixed to such a value so that the model can reproduce the late time accelerated expansion of the universe. The minus sign in $(-T)^n$ has also be chosen with some foresight, as $T=-\frac{2}{3}\theta^2=-6H^2<0$ in background cosmology. Our aim is to examine if this particular class of model which can reproduce the background cosmological evolution of the $\Lambda$CDM model is also compatible with large scale structure evolution.
Such a Lagrangian has been studied previously by
\cite{Linder:2010, Zheng:2010} but in different contexts.  

\subsection{Background Evolution}

In background cosmology, the modified Friedman equation is given as
\begin{eqnarray}
3H^2 &=& \kappa\left(\rho^f+\rho^{eff}\right)\nonumber\\
&=& \frac{1}{f_T}\kappa\rho^f - \frac{1}{2f_T}\left(f-f_TT\right).
\end{eqnarray}
Using the fact that $T=-6H^2$, this equation could be written as
\begin{eqnarray}
-T-(1+2n)\frac{\mu^{2(n+1)}}{(-T)^n} &=& 2\kappa\rho^f,
\end{eqnarray}
according to which we could fix $\mu$ by assuming that the present
fractional energy density for ``dark energy" is $\Omega_\Lambda$:
\begin{eqnarray}
\mu^{2(1+n)} &=& \frac{1}{1+2n}\Omega_\Lambda\left(6H_0^2\right)^{1+n}.
\end{eqnarray}
Here $H_0$ is the present Hubble expansion rate.  The
modified Friedman equation can then take the form
\begin{eqnarray}\label{eq:modified_friedman}
3H^2 &=& \kappa\rho^f+3\Omega_\Lambda H_0^2\left(\frac{H_0}{H}\right)^n.
\end{eqnarray}
Here the second term in the right-hand side represents the energy
density of an effective dark energy component. Given the value for
$n$, we can solve the algebraic equation
Eq.~(\ref{eq:modified_friedman}) to find the expansion rate of the
Universe at any earlier time.

We have considered five different values for $n$, with $n=0.0,\pm 0.1,\pm 0.2
$, and summarised the results for the background evolution in Fig.~\ref%
{fig:background}. The upper left panel shows the fractional energy densities
for matter, radiation and effective dark energy respectively. The black
solid curve ($n=0$) is the $\Lambda $CDM paradigm. 
$H_0/H$ increases until
it reaches its current value $1.0$, so a positive $n$ (green dotted
and purple dash-dotted curves; same below) means the  energy
density of dark energy was lower in the past. The opposite is true for a  negative $n$ (cyan dashed and
pink dash-triple-dotted curves; same below). This behaviour is as predicted by 
Eq.~(\ref{eq:modified_friedman}).

For positive values of $n$ the energy density of the ``dark energy" 
increases in time, which implies that its pressure-density ratio should be
less than $-1$. Given that we normalise the ``dark energy" fractional energy density by its value today, at earlier times it will be lower in the $f(T)$ gravity model than in $\Lambda$CDM and so the universe will expands slower than in the latter. The effect on the total effective pressure-density ratio of all matter species, which is defined as $w_{eff}\equiv=-1-\frac{2\dot{H}}{3H^2}$, is shown in
the upper-right panel of Fig.~\ref{fig:background}.

Meanwhile, since for positive values of $n$ the dark energy (and
therefore the total energy) density was lower in the past than in
the $\Lambda$CDM paradigm, the Hubble expansion rate for the
former must be lower too, as can be seen from the lower-left panel
of Fig.~\ref{fig:background}. Note that at very early times the
expansion rates in these two models are almost the same, because
the effect of the $f(T)$ correction (or the cosmological constant)
is negligible then.

Finally, we shall find that the quantity $f_T=df/dT$ is important
in the $f(T)$ gravity model and so have plotted its evolution in
the lower-right panel of Fig.~\ref{fig:background}. Clearly
\begin{eqnarray}
f_T &=& 1-n\frac{\mu^{2(n+1)}}{(-T)^{n+1}}
\end{eqnarray}
and therefore must be negative for positive values of $n$,
and vice versa. Again, at very early times 
$f_T-1\approx0$ because
$|f(T)-1|\ll|T|$, and the deviation of $f_{T}$ from unity
only becomes large at late times.

These results show that as long as $|n|$ is close enough to 
$0$, the deviation of the $f(T)$ gravity model from $\Lambda $CDM is small
but the background expansion rate could provide a weak
constraint on the model parameter $n$.

\subsection{CMB and Large-scale Structure}

Having fixed $\mu$ in order to reproduce the desired background evolution, we are ready to consider the evolution of linear
perturbations. These could place much more stringent constraints on the
model parameters.

In the section above we gave the covariant and gauge invariant
linear perturbation equations for general $f(T)$ models. In order to solve these equation numerically  we must specify a gauge (or reference frame). As
usual, we choose to work in the CDM frame (that is, the reference frame of an observer
comoving with dark matter fluid), which is characterised by
$v_{\mathrm{CDM}}=A=0$, where $v_{\mathrm{CDM}}$ is the peculiar
velocity of the dark matter fluid and $A$ is the acceleration of
the observer.

Next, we need to determine the behaviour of the new degree of freedom
$\ae$. In most modified gravity theories, this will be governed by a
dynamical equation. In the $f(T)$ gravity, however, its value is given by a
constraint equation. 
This is a consequence of the fact that the right-hand side of Eq.~(\ref{eq:field_eqn}) is not {\em a priori} antisymmetric, but it is required to be as a consequence of the field equations. This leads to the two different expressions in 
Eqs.~(\ref{eq:q_tot_a}) and (\ref{eq:q_tot_b}), which imply that 
\begin{eqnarray}
k\mathcal{H}(\mathcal{Z}+\ae) &=& \frac{3}{2}\left(\mathcal{H}'-\mathcal{H}^2\right)\eta\,.
\end{eqnarray}
This equation can be used to determine $\ae$ in terms of $\mathcal{Z}, \eta$ and background
quantities.

We can then eliminate $\ae$ in all the relevant
perturbation equations. Nonetheless, it is interesting to see how the new degree 
of freedom $\ae$ evolves in time on different length scales, and this
is shown in Fig.~\ref{fig:ae}. Since $\ae$ is not a gauge invariant quantity,
what we have plotted is $\epsilon\equiv\ae+\sigma$, which is gauge invariant. Note that
in the conformal Newtonian gauge, in which
$\sigma=0$ (c.f.~Appendix~\ref{appen:newtonian}), the quantity $\epsilon$ coincides with $\ae$. We show
the results for $n=0.1$ in Fig.~\ref{fig:ae}. We see that $\epsilon$ decreases
in time, and the decrease becomes more rapid as one moves to smaller scales 
(bigger $k$'s). Therefore, we expect any deviations from the $\Lambda$CDM model to be more important on large scales than
on small scales. We will confirm this below.

We can now examine the growth of the dark-matter density contrast in the
context of the $f(T)$ gravity model. For simplicity, we shall assume that
the universe is filled with dark matter only, which is a fair approximation
at late times. 
Taking the spatial derivative of
the Raychaudhuri equation one gets \cite{Li:2009}
\begin{eqnarray}
\label{interim}
&&k\mathcal{Z}' + k\mathcal{HZ} -k^2A\\&&\qquad\qquad + 3\left(\mathcal{H}'-\mathcal{H}^2\right)A = -\frac{1}{2}\left(\mathcal{X}+3\mathcal{X}^p\right)a^2,\nonumber
\end{eqnarray}
in which $k$ is the wavenumber and $\mathcal{X}, \mathcal{X}^p$
include contributions from both the dark matter and the $f(T)$
corrections. In the CDM frame $A=0$, and the conservation
equation for dark matter gives $\Delta'=-k\mathcal{Z}$, where
$\Delta=\mathcal{X}_{\mathrm{DM}}/\rho_{\mathrm{DM}}$ is the dark
matter density contrast. Then, Eq.~(\ref{interim}) can be rewritten, by
manipulating our set of perturbation equations, as
\begin{eqnarray}\label{eq:delta_dm}
\Delta''+\left(1-2C\right)H\Delta' &=& \frac{\kappa\rho_{\mathrm{DM}}a^2}{f_T}\left[\frac{1}{2}+C\right]\Delta
\end{eqnarray}
with $C$ defined by
\begin{eqnarray}
C &\equiv& 216\frac{f_{TTT}/a^4}{f_T}\frac{\mathcal{H}^2\left(\mathcal{H}'-\mathcal{H}^2\right)^2}{k^2-36\frac{f_{TT}/a^2}{f_T}\mathcal{H}^2\left(\mathcal{H}'-\mathcal{H}^2\right)}\nonumber\\
&&-216\frac{\left[\frac{f_{TT}/a^2}{f_T}\mathcal{H}\left(\mathcal{H}'-\mathcal{H}^2\right)^2\right]^2}{k^2-36\frac{f_{TT}/a^2}{f_T}\mathcal{H}^2\left(\mathcal{H}'-\mathcal{H}^2\right)}\nonumber\\
&&+\frac{f_{TT}/a^2}{f_T}\frac{156\mathcal{H}^2\mathcal{H}'-24\mathcal{H}\mathcal{H}''-60\mathcal{H}^3-48\mathcal{H}'^2}{k^2-36\frac{f_{TT}/a^2}{f_T}\mathcal{H}^2\left(\mathcal{H}'-\mathcal{H}^2\right)}.\nonumber
\end{eqnarray}
Clearly, on very small scales, where $k\gg \mathcal{H},%
\mathcal{H}^{\prime }/\mathcal{H}$ and $\mathcal{H}^{\prime \prime }/%
\mathcal{H}^{2}$ we have $C\rightarrow 0$ and Eq.~(\ref{eq:delta_dm})
reduces to that in the $\Lambda$CDM model, only with the
value of the gravitational constant rescaled by $1/f_T$.
On very large scales, in contrast, we can neglect $k^{2}$ in the expression
for $C$, and Eq.~(\ref{eq:delta_dm}) becomes very complicated, leading to
large deviations from $\Lambda $CDM.

One should also be able to derive an evolution equation for the %
gravitational potential $\phi$ defined in
Eq.~(\ref{eq:HarmonicExpansion}) (indeed this will be easier if we use the
Newtonian gauge potentials given in Appendix~\ref{appen:newtonian}), but we
shall not do that here.

In Fig.~\ref{fig:linpert} we show some results for the linear perturbation
evolutions in the $f(T)$ model studied here. Clearly both the CMB and matter
power spectra (for all choices of $n$ except for $n=0$ which corresponds to $%
\Lambda $CDM) blow up on
large angular scales (small $\ell$ or small $k$), which is
consistent with the above analysis that the evolution of matter
density perturbations (and therefore the gravitational potential)
on large scales is very different from the $\Lambda$CDM
predictions. On small scales, however, the $f(T)$ model gives
similar predictions as $\Lambda$CDM, which is as expected.

To see more clearly how the growth of the dark matter
density contrast and the growth of the gravitational
potential have been modified, we have plotted them in the
lower panels of Fig.~\ref{fig:linpert}. For $\Delta _{\mathrm{CDM}}$, the
difference between the $f(T)$ models (with $n=\pm 0.1$) and the $\Lambda $CDM is within $\sim 10\%$ on small scales ($k>0.001h$~Mpc$^{-1}$) because
the effective gravitational constant is rescaled and the cosmic expansion
rate is modified as well. But on very large scales ($k<0.0001h$~Mpc$^{-1}$), the difference becomes very significant. 
The same happens to $\phi$.

These results are expected to remain qualitatively true for other choices
for the function $f(T)$, if they are made so as to explain the late-time
acceleration of the universe. This can be seen from the expression 
for $C$, which shows that the
large-scale deviation from $\Lambda$CDM is inevitable whenever
$f_{TT}$ and/or $f_{TTT}$ are nonzero.

The results suggest that $f(T)$ gravity models which are proposed as an alternative to
dark energy could face severe difficulties in being compatible with observations regarding large scale evolution. The expectation that that linear perturbation
analysis gives better constraints than the consideration of
background cosmology alone is clearly confirmed here as well.

\section{Summary and Conclusions}

\label{sect:con}

In summary, we have given the modified Einstein equations for general $f(T)$
gravity models in a covariant formalism, and derived the covariant and
gauge-invariant perturbation equations in the $3+1$ formalism. The
perturbation equations take full account of the extra degrees of freedom in
the $f(T)$ gravity theory (the importance of which was first discussed in 
Ref.~\cite{Li:2010}) up to linear order. The equations in specific gauges can
then be obtained straightforwardly as shown in Appendix~\ref{appen:newtonian}.

For a general $f(T)$ theory it turns out that
no new degrees of freedom 
appear at the background level, and the modified Friedmann equation
is simply a nonlinear algebraic equation in the Hubble rate $H$ that can easily be solved numerically.
At the linear order in perturbation there is a new vector degree of freedom (as a consequence of the lack of
local Lorentz symmetry, as pointed out in Ref.~\cite{Li:2010}). However, at this order the equations include no time derivatives of this vector, which just satisfies a constraint equation.

After developing the general formalism and deriving the perturbed equation at linear order, we restricted our attention to scalar perturbations. We then considered a broad class of $f(T)$ theories which are representative examples of models that could account for the late-time acceleration of the universe, as proposed in the literature. We studied in detail their background
cosmology and the evolution of linear perturbations. We were able to determine the new degree of freedom algebraically  in
terms of other curvature perturbation quantities. 
We also derived the evolution equation for the dark-matter density
contrast $\Delta$ in a dark-matter-dominated universe, and showed
that it resembles that of $\Lambda$CDM on small scales, but gets
significantly modified on large scales. The
large-scale CMB and matter power spectra blow up, signalling a serious viability problem for any
 $f(T)$ models that are able to account for the accelerated
expansion of the universe at the background level. We have argued that this conclusion is
robust and holds true for other choices of $f(T)$ unless
$f_{TT}=f_{TTT}=0$ at late times.

Our result clarifies the effects of the new degree of freedom in the $f(T)$ gravity model
at the linear perturbation level, and we have seen here that only one extra degree of freedom arises. An interesting
question is whether further degrees of freedom will enter into the field equations, and if so, whether they are well behaved, when followed beyond
linear perturbation. This will be investigated elsewhere.

\appendix

\section{$T$ and $S_{\nu\mu\rho}\nabla^{\rho}T$ up to First Order}

\label{appen:expression}

We give the perturbative expansions and calculations needed to derive Eqs.~(\ref{perT}) and (\ref{perS}). First,  we need to express the covariant
derivatives of $h_{\mu }^{\underline{i}}$ and $\ae _{\mu }$ in terms of
perturbation quantities in the $3+1$ formalism. Using the definition $\hat{\nabla}_{\mu }h_{\mu }^{\underline{i}}=H_{\mu }^{\alpha }H_{\nu }^{\beta
}\nabla _{\alpha }h_{\beta }^{\underline{i}}$ it is straightforward to show 
\begin{eqnarray}\label{eq:decomp3}
\nabla_{\mu}h^{\underline{i}}_{\nu} &\approx&
u_{\mu}\dot{h}^{\underline{i}}_{\nu} + \hat{\nabla}_\mu
h^{\underline{i}}_\nu + \frac{1}{3}\theta u_\mu u_\nu
U^{\underline{i}} +
u_\nu\hat{\nabla}_{\mu}U^{\underline{i}}\nonumber\\
&&-u_\nu\left(\frac{1}{3}\theta
h^{\underline{i}}_{\mu}+h^{\underline{i}}_{\beta}\sigma_{\mu}^{\
\beta}+h^{\underline{i}}_{\beta}\varpi_{\mu}^{\ \beta}\right)
\end{eqnarray}
up to first order. Note that $\dot{h}^{\underline{i}}_{\nu}$ and
$\hat{\nabla}_\mu h^{\underline{i}}_\nu$ are both first order.
Similarly, for $\ae_{\mu}$, which is itself first
order, we have
\begin{eqnarray}\label{eq:decomp4}
\nabla_\mu\ae_\nu &\approx& u_{\mu}\dot{\ae}_{\nu} +
\hat{\nabla}_{\mu}\ae_{\nu} - \frac{1}{3}\theta u_{\nu}\ae_{\mu}.
\end{eqnarray}

Next we consider $T$. Using Eqs.~(\ref{eq:K_cov}) and (\ref{eq:Ricci}) we
find that
\begin{eqnarray}
T &=& K^{\mu\nu\rho}K_{\rho\nu\mu} - K^{\mu\rho}_{\ \
\mu}K^{\nu}_{\ \rho\nu}\nonumber\\
&=&
\left(\nabla^{\nu}h^{\mu}_{a}\right)\left(\nabla_{\mu}h^{a}_{\nu}\right)
-\eta_{ab}\left(\nabla_{\mu}h^{\mu}_{a}\right)\left(\nabla_{\nu}h^{\nu}_{a}\right).
\end{eqnarray}
Then, given Eq.~(\ref{eq:decomp3}), we can show that
\begin{eqnarray}
\left(\nabla^{\nu}h^{\mu}_{\underline{i}}\right)\left(\nabla_{\mu}h^{\underline{i}}_{\nu}\right)\
\approx\
\eta^{\underline{i}\underline{j}}\left(\nabla_{\mu}h^{\mu}_{\underline{i}}\right)\left(\nabla_{\nu}h^{\nu}_{\underline{j}}\right)\
\approx\ 0\nonumber
\end{eqnarray}
to first order, and therefore
\begin{eqnarray}\label{eq:T_pert}
T &\approx&
\left(\nabla^{\nu}h^{\mu}_{\underline{0}}\right)\left(\nabla_{\mu}h^{\underline{0}}_{\nu}\right)
-\eta^{\underline{0}\underline{0}}\left(\nabla_{\mu}h^{\mu}_{\underline{0}}\right)\left(\nabla_{\nu}h^{\nu}_{\underline{0}}\right)\nonumber\\
&\approx&
-\frac{2}{3}\theta^2-\frac{4}{3}\theta\hat{\nabla}^{\mu}\ae_{\mu}.
\end{eqnarray}

Similarly, it can be shown that
\begin{eqnarray}
S_{\nu\mu\rho}\nabla^{\rho}T &=& u^{\rho}\dot{T}S_{\nu\mu\rho} +
S_{\nu\mu\rho}\hat{\nabla}^{\rho}T
\end{eqnarray}
where, to first order,
\begin{eqnarray}
S_{\nu\mu\rho}\hat{\nabla}^{\rho}T &\approx& -\frac{2}{3}\theta
u_{\mu}\hat{\nabla}_{\nu}T.
\end{eqnarray}
According to Eq.~(\ref{eq:S_cov}),
\begin{eqnarray}
u^{\rho}S_{\nu\mu\rho} &=&
u_{\rho}h^{a}_{\mu}\nabla_{\nu}h^{\rho}_{a} +
u_{\nu}h^{\lambda}_{a}\nabla_{\lambda}h^{a}_{\mu} -
g_{\mu\nu}u^{\rho}h_{a}^{\lambda}\nabla_{\lambda}h^{a}_{\rho}\nonumber
\end{eqnarray}
with, to the same order,
\begin{eqnarray}
u_{\rho}h^{a}_{\mu}\nabla_{\nu}h^{\rho}_{a} &\approx& -
\frac{1}{3}\left(\theta+\hat{\nabla}^{\rho}\ae_{\rho}\right)H_{\mu\nu}
- u_{\nu}\left(A_\mu+\dot{\ae}_{\mu}\right)\nonumber\\
&&-\left(\sigma_{\mu\nu}+\varpi_{\mu\nu}+\hat{\nabla}_{\langle\mu}\ae_{\nu\rangle}
+\hat{\nabla}_{[\nu}\ae_{\mu]}\right),\ \ \ \nonumber\\
- g_{\mu\nu}u^{\rho}h_{a}^{\lambda}\nabla_{\lambda}h^{a}_{\rho}
&\approx&
\left(\theta+\hat{\nabla}^{\rho}\ae_{\rho}\right)g_{\mu\nu},\nonumber\\
\label{eq:use3}u_{\nu}h^{\lambda}_{a}\nabla_{\lambda}h^{a}_{\mu}
&\approx& -\left(\theta+\hat{\nabla}^{\rho}\ae_{\rho}\right)u_\mu
u_\nu
-\theta u_\nu\ae_\mu\nonumber\\
&&-u_\nu\left(h^{\underline{i}}_{\mu}\nabla_{\lambda}h^{\lambda}_{\underline{i}}\right).
\end{eqnarray}
Clearly now we need to calculate
$\left(h^{\underline{i}}_{\mu}\nabla_{\lambda}h^{\lambda}_{\underline{i}}\right)$.
Note that this is a vector which is first order in perturbation,
and
$u^{\mu}\left(h^{\underline{i}}_{\mu}\nabla_{\lambda}h^{\lambda}_{\underline{i}}\right)
=U^{\underline{i}}\left(\nabla_{\lambda}h^{\lambda}_{\underline{i}}\right)\approx0$,
which means that the part of
$\left(h^{\underline{i}}_{\mu}\nabla_{\lambda}h^{\lambda}_{\underline{i}}\right)$
which is parallel to $u^{\mu}$ vanishes up to first order, so we need to consider only the part perpendicular to $u^{\mu}$,
$\left(h^{\underline{i}}_{\mu}\nabla_{\lambda}h^{\lambda}_{\underline{i}}\right)_{\perp}\equiv\Upsilon_{\mu}$.

In order to find an expression for $\Upsilon_\mu$, consider
Eq.~(\ref{eq:Ricci}), $T^{\nu\mu}_{\ \ \mu}=K^{\mu\nu}_{\ \ \nu}$
and Eq.~(\ref{eq:K_cov}), which leads to
\begin{eqnarray}
2\nabla^{\mu}\left(h^{a}_{\mu}\nabla_\nu h^{\nu}_{a}\right) &=&
-R-T.
\end{eqnarray}
Using Eq.~(\ref{eq:T_pert}) and the relation
\begin{eqnarray}
R &\approx& -2\dot{\theta} - \frac{4}{3}\theta^2 +
2\hat{\nabla}^{\mu}A_{\mu} - \hat{R}
\end{eqnarray}
to first order \cite{Li:mgb}, we have that
\begin{eqnarray}\label{eq:use1}
2\nabla^{\mu}\left(h^{a}_{\mu}\nabla_\nu h^{\nu}_{a}\right)
&\approx& 2\dot{\theta}+2\theta^2-2\hat{\nabla}\cdot A\nonumber\\
&&+\hat{R}+\frac{4}{3}\theta\hat{\nabla}\cdot\ae.
\end{eqnarray}
On the other hand, writing
\begin{eqnarray}
h^{a}_{\mu}\nabla_\nu h^{\nu}_{a} &=&
h^{\underline{0}}_{\mu}\nabla_\nu
h^{\nu}_{\underline{0}}+h^{\underline{i}}_{\mu}\nabla_\nu
h^{\nu}_{\underline{i}}\nonumber\\
&=& h^{\underline{0}}_{\mu}\nabla_\nu
h^{\nu}_{\underline{0}}+\Upsilon_\mu\nonumber
\end{eqnarray}
it is easy to obtain
\begin{eqnarray}\label{eq:use2}
2\nabla^{\mu}\left(h^{a}_{\mu}\nabla_\nu h^{\nu}_{a}\right)
&\approx&
2\dot{\theta}+2\theta^2+2\hat{\nabla}^{\mu}\dot{\ae}_{\mu}\nonumber\\
&&+ \frac{10}{3}\theta\hat{\nabla}^{\mu}\ae_{\mu} +
2\hat{\nabla}^{\mu}\Upsilon_{\mu}
\end{eqnarray}
where we have used
$\nabla^{\mu}\Upsilon_{\mu}\approx \hat{\nabla}^{\mu}\Upsilon_{\mu}$
because $\Upsilon_\mu$ is first order. From
Eqs.~(\ref{eq:use1}, \ref{eq:use2}) we have
\begin{eqnarray}
\hat{\nabla}^{\mu}\Upsilon_{\mu} &\approx&
-\hat{\nabla}^{\mu}\dot{\ae}_{\mu}-\theta\hat{\nabla}^{\mu}\ae_{\mu}-\hat{\nabla}^{\mu}A_{\mu}+\frac{1}{2}\hat{R}\,.\nonumber
\end{eqnarray}
For a sufficiently well behaved $\hat{R}$, which is the case here, one can write $\hat{\nabla}^{\mu}\hat{R}_{\mu}=\hat{R}$, where $\hat{R}_{\mu}$ is actually the gradient of a scalar. Then
\begin{eqnarray}
\Upsilon_{\mu} &\approx&
-\dot{\ae}_{\mu}-\theta\ae_{\mu}-A_{\mu}+\frac{1}{2}\hat{R}_{\mu}\,.
\end{eqnarray}
Substituting this
back into Eq.~(\ref{eq:use3}), we get
\begin{eqnarray}
S_{\nu\mu\rho}\nabla^{\rho}T &\approx& \frac{2}{3}\dot{T}\left(\theta+\hat{\nabla}\cdot\ae\right)H_{\mu\nu}-\frac{1}{2}\dot{T}u_{\nu}\hat{R}_{\mu}\nonumber\\
&&- \dot{T}\left(\sigma_{\mu\nu}+\varpi_{\mu\nu}+\hat{\nabla}_{\langle\mu}\ae_{\nu\rangle}+\hat{\nabla}_{[\mu}\ae_{\nu]}\right)\nonumber\\
&&-\frac{2}{3}\theta u_\mu\hat{\nabla}_{\nu}T.
\end{eqnarray}

\section{Perturbation Equations in the Newtonian Gauge}

\label{appen:newtonian}

The conformal Newtonian gauge can be obtained by setting $\sigma=0$. Defining
\begin{eqnarray}
\Psi &\equiv& \phi-\frac{\kappa\Pi a^2}{2k^2},\nonumber\\
\Phi &\equiv& \phi+\frac{\kappa\Pi a^2}{2k^2},
\end{eqnarray}
and manipulating
Eqs.~(\ref{eq:constraint_q2}) to (\ref{eq:propagation_eta2}), we obtain
\begin{eqnarray}
A &=& -\Psi,\nonumber\\
k\mathcal{Z} &=& -3\left(\Phi'+\mathcal{H}\Psi\right),\nonumber\\
\eta &=& -2\Phi.
\end{eqnarray}
With these, and using Eqs.~(\ref{eq:constraint_q2}) to
(\ref{eq:propagation_eta2}) and (\ref{eq:delta_rho_tot}) to 
(\ref{eq:pi_tot}), the perturbed field equations in the
Newtonian gauge are derived as
\begin{widetext}
\begin{eqnarray}
\frac{1}{2}\kappa\delta\rho^{f}a^2 &=& -f_Tk^2\Phi
-
3\mathcal{H}\left(\Phi'+\mathcal{H}\Psi\right)\left[f_T-12\frac{f_{TT}}{a^2}\mathcal{H}^2\right]
-12\frac{f_{TT}}{a^2}k\mathcal{H}^{3}\ae,\\
\frac{1}{2}\kappa\delta p^{f}a^2 &=&
f_T\left[\Phi''+\mathcal{H}\left(\Psi'+2\Phi'\right)+
\left(2\mathcal{H}'+\mathcal{H}^2\right)\Psi+\frac{1}{3}k^2(\Phi-\Psi)\right]
-48\frac{f_{TTT}}{a^4}k\mathcal{H}^3\left(\mathcal{H}'-\mathcal{H}^2\right)\ae\nonumber\\
&&-12\frac{f_{TT}}{a^2}\left[\mathcal{H}^2\Phi''+\mathcal{H}\left(3\mathcal{H}'-
\mathcal{H}^2\right)\Phi'+\mathcal{H}^3\Psi'+\left(5\mathcal{H}'-2\mathcal{H}^2\right)\mathcal{H}^2\Psi\right]\nonumber\\
&&+144\frac{f_{TTT}}{a^4}\mathcal{H}^3\left(\mathcal{H}'-\mathcal{H}^2\right)\left(\Phi'+\mathcal{H}\Psi\right)
+4\frac{f_{TT}}{a^2}\left[k^{2}\mathcal{H}\ae+k\mathcal{H}\left(3\mathcal{H}'-\mathcal{H}^2\right)\ae\right],\\
\frac{1}{2}\kappa q^{f}a^2 &=&
\left(f_T-12\frac{f_{TT}}{a^2}\mathcal{H}^2\right)k\left(\Phi'+\mathcal{H}\Psi\right)
+4\frac{f_{TT}}{a^2}k^2\mathcal{H}^2\ae\\
&=& f_Tk(\Phi'+\mathcal{H}\Psi)
-12\frac{f_{TT}}{a^2}k\mathcal{H}\left(\mathcal{H}'-\mathcal{H}^2\right)\Phi,\\
\kappa\Pi^{f}a^2 &=& f_Tk^2(\Phi-\Psi)
+12\frac{f_{TT}}{a^2}k\mathcal{H}\left(\mathcal{H}'-\mathcal{H}^2\right)\ae.
\end{eqnarray}
\end{widetext}
Obviously when $f_T-1=f_{TT}=f_{TTT}=0$ these equations reduce to
those of general relativity.

\

\begin{acknowledgments}
BL is supported by Queens' College, University of Cambridge and
the rolling grant from Science and Technology Facilities Council
({\tt STFC}) of the UK. TPS is supported by a Marie Curie Fellowship. The linear
perturbation computation in this work has been performed using a
modified version of the publicly-available {\tt CAMB} code
\cite{CAMB}.
\end{acknowledgments}


\begin{thebibliography}
\bibitem{} \ifx\csname natexlab\endcsname\relax \fi \expandafter\ifx\csname
bibnamefont\endcsname\relax

\fi \expandafter\ifx\csname bibfnamefont\endcsname\relax

\fi \expandafter\ifx\csname citenamefont\endcsname\relax

\fi \expandafter\ifx\csname url\endcsname\relax

\fi \expandafter\ifx\csname urlprefix\endcsname\relax

\fi \providecommand{\bibinfo}[2]{#2} \providecommand{\eprint}[2][]{\url{#2}}

\bibitem[Bengochea \& Ferraro (2009)]{Bengochea:2009} G.~R.~Bengochea and R.~Ferraro, Phys.~Rev.~D{\bf79}, 124019 (2009).

\bibitem[Yu \& Wu (2010a)]{Yu:2010a} P.~Wu and H.~Yu (2010), arXiv:1006.0674 [gr-qc].

\bibitem[Myrzakulov (2010a)]{Myrzakulov:2010a} R.~Myrzakulov (2010), arXiv:1006.1120 [astro-ph.CO].

\bibitem[Tsyba {\it et~al.}(2010)]{Tsyba:2010} P.~Yu~Tsyba, I.~I.~Kulnazarov, K.~K.~Yerzhanov and R.~Myrzakulov (2010), arXiv:1008.0779 [astro-ph.CO].

\bibitem[Linder (2010)]{Linder:2010} E.~V.~Linder, Phys.~Rev.~D{\bf81}, 127301 (2010).

\bibitem[Yu \& Wu (2010b)]{Yi:2010b} P.~Wu and H.~Yu (2010), arXiv:1008.3669 [astro-ph.CO].

\bibitem[Kazuharu, Geng \& Lee (2010a)]{Kazuharu:2010a} K.~Bamba, C.~Q.~Geng and C.~C.~Lee (2010), arXiv:1008.4036 [astro-ph.CO].

\bibitem[Kazuharu, Geng \& Lee (2010b)]{Kazuharu:2010b} K.~Bamba, C.~Q.~Geng and C.~C.~Lee, JCAP, {\bf08}, 021 (2010).

\bibitem[Myrzakulov (2010b)]{Myrzakulov:2010b} R.~Myrzakulov (2010), arXiv:1008.4486 [astro-ph.CO].

\bibitem[Yu \& Wu (2010c)]{Yu:2010c} P.~Wu and H.~Yu, Phys.~Lett.~B{\bf692}, 176 (2010).

\bibitem[Karami \& Abdolmaleki (2010)]{Karami:2010} K.~Karami and A.~Abdolmaleki (2010), arXiv: 1009.2459 [gr-qc].

\bibitem[Dent, Dutta \& Saridakis (2010)]{Dent:2010} S.~-H.~Chen, J.~B.~Dent, S.~Dutta and E.~N.~Saridakis, Phys.~Rev.~D{\bf 83}, 023508 (2011).

\bibitem[Li, Sotiriou \& Barrow (2010)]{Li:2010} B.~Li, T.~P.~Sotiriou and J.~D.~Barrow, Phys.~Rev.~D, in press; arXiv:1010.1041 [gr-qc].

\bibitem[Zheng \& Huang (2010)]{Zheng:2010} R.~Zheng and Q.~Huang (2010), arXiv:1010.3512 [astro-ph.CO].

\bibitem[Dent, Dutta \& Saridakis (2011)]{Dent:2011} J.~B.~Dent, S.~Dutta and E.~N.~Saridakis, J.~Cosmo.~Astropart.~Phys., {\bf01}, 009 (2011).

\bibitem[Sotiriou, Li \& Barrow (2010)]{Sotiriou:2010} T.~P.~Sotiriou, B.~Li and J.~D.~Barrow (2010), arXiv:1012.4039 [gr-qc].

\bibitem[Zhang {\it et~al.} (2011)]{Zhang:2011} Y.~Zhang, H.~Li, Y.~Gong and Z.~Zhu (2011), arXiv:1103.0719 [astro-ph.CO].

\bibitem[Unzicker \& Case (2005)]{Einstein} A.~Unzicker and T.~Case (2005), arXiv:physics/0503046.

\bibitem[Aldrovandi \& Pereira (2010)]{Aldrovandi} R.~Aldrovandi and J.~G.~Pereira, {\it An Introduction to Teleparallel Gravity}, Instituto de Fisica Teorica, UNSEP, Sao Paulo (http://www.ift.unesp.br/gcg/tele.pdf).

\bibitem[Weitzenbock (1923)]{Weitzenbock} R.~Weitzenbock, {\it Invariance Theorie}, Nordhoff, Groningen, 1923.

\bibitem{Sotiriou:2008rp} T.~P.~Sotiriou and V.~Faraoni, Rev.\ Mod.\ Phys.\  {\bf 82}, 451 (2010).

\bibitem[Carroll {\it et~al.}(2005)]{Carroll:2005} S.~M.~Carroll, A.~De~Felice, V.~Duvvrui, D.~A.~Easson, M.~Trodden and M.~S.~Turner, Phys.~Rev.~D{\bf71}, 063513 (2005).

\bibitem[Weinberg (1972)]{weinbergbook} S.~Weinberg, {\it Gravitation and Cosmology: Principles and Applications of the General Theory of Relativity}, John Wiley \& Sons, New York, 1972.

\bibitem[Ferraro \& Fiorini (2007)]{Ferraro:2007} R.~Ferraro and F.~Fiorini, Phys.~Rev.~D{\bf75}, 084031 (2007).

\bibitem[Ellis {\it et~al.}(1989)]{GR3+1} G.~F.~R.~Ellis and H.~Van~Elst, in \emph{Theoretial and Observational Cosmology,} edited by M. Lachi\`{e}ze-Rey (Springer, New York, 1998);
A.~Challinor and A.~Lasenby, Astrophys.~J. \textbf{513}, 1 (1999).

\bibitem[Li \& Chu (2006)]{Li:pfr-1} B.~Li and M.~-C.~Chu, Phys.~Rev.~D{\bf74}, 104010 (2006).

\bibitem[Li, Chan \& Chu (2007)]{Li:pfr-2} B.~Li, K.~C.~Chan and M.~-C.~Chu, Phys.~Rev.~D{\bf76}, 024002 (2007).

\bibitem[Li \& Barrow (2007)]{Li:mfr} B.~Li and J.~D.~Barrow, Phys.~Rev.~D{\bf75}, 084010 (2007).

\bibitem[Li, Barrow \& Mota (2007)]{Li:mgb} B.~Li, J.~D.~Barrow and D.~F.~Mota, Phys.~Rev.~D{\bf76}, 044027 (2007)

\bibitem[Li, Mota \& Barrow (2008)]{Li:aether} B.~Li, D.~F.~Mota and J.~D.~Barrow, Phys.~Rev.~D{\bf77}, 024032 (2008).

\bibitem{Ferraro:2011us} R.~Ferraro and F.~Fiorini, arXiv:1103.0824 [gr-qc].

\bibitem[Li \& Barrow (2009)]{Li:2009} B.~Li and J.~D.~Barrow, Phys.~Rev.~D{\bf79}, 103521 (2009).

\bibitem[Lewis, Challinor \& Lasenby (2000)]{CAMB} A.~M.~Lewis, A.~Challinor and A.~N.~Lasenby, Astrophys.~J.~{\bf538}, 473 (2000).

\end{thebibliography}
\end{document}